\documentclass{ws-ijqi}

\begin{document}

\markboth{G. Adenier et al.}
{Realization of a twin beam source based on four-wave mixing in Cesium}

\catchline{}{}{}{}{}

\title{Realization of a twin beam source based on four-wave mixing in Cesium}

\author{G. Adenier$^{1}$, D. Calonico$^{1}$, S. Micalizio$^{1}$, N. Samantaray$^{1,2}$, I. P. Degiovanni$^{1}$ and I.~Ruo~Berchera$^{1}$ }

\address{$^{1}$INRiM, strada delle Cacce 91,\\10135, Torino, Italy}

\address{$^{2}$Politecnico di Torino, Corso Duca degli Abruzzi, 24\\10129 Torino, Italy}

\maketitle

\begin{history}
\received{Day Month Year}
\revised{Day Month Year}
\end{history}

\begin{abstract}
Four-wave mixing (4WM) is a known source of intense non-classical twin beams. It can be generated when an intense laser beam (the pump) and a weak laser beam (the seed) overlap in a $\chi^{(3)}$ medium (here cesium vapor), with frequencies close to resonance with atomic transitions. The twin beams generated by 4WM have frequencies naturally close to atomic transitions, and can be intense (gain $\gg 1$) even in the CW pump regime, which is not the case for PDC $\chi^{(2)}$ phenomenon in non-linear crystals.  So, 4WM is well suited for atom-light interaction and atom-based quantum protocols. Here we present the first realization of a source of 4-wave mixing exploiting $D_2$ line of Cesium atoms.
\end{abstract}


\section{Introduction}	
In the last years four-wave mixing (4WM) has become attractive as a high gain source of intense non classical twin beams in the CW pump regime. Twin beams are generated by interaction in a $\chi^{(3)}$ medium (here Cesium vapor), with frequencies close to resonance with atomic transitions. So, 4WM is well suited for atom-light interaction and atom-based quantum protocols. It has been studied experimentally extensively in the literature, in particular with Rubidium (see Ref.~[1] and references therein), and very recently with the $D_1$ line of Cesium atoms (see Ref.~[2]), thereby exhibiting intensity correlated beams with noise reduction below the standard quantum limit (SQL). Nevertheless, 4WM is considered difficult to achieve with Cesium atoms because of the large hyperfine splitting of the ground state, in particular with diode lasers due to high phase noise (see Ref.~[2]). Based on the results that we present here on the $D_2$ line of Cesium atoms, we demonstrate that these difficulties can be overcome and a twin beam source around that wavelength ($\lambda=$852nm) can be generated. The application are in the field of quantum technologies, concerning non-classical correlations and photon number entanglement [3-26] (for a recent review see Ref.~[27]).

\section{Theory}
\begin{figure}
\centering \includegraphics[width=5cm]{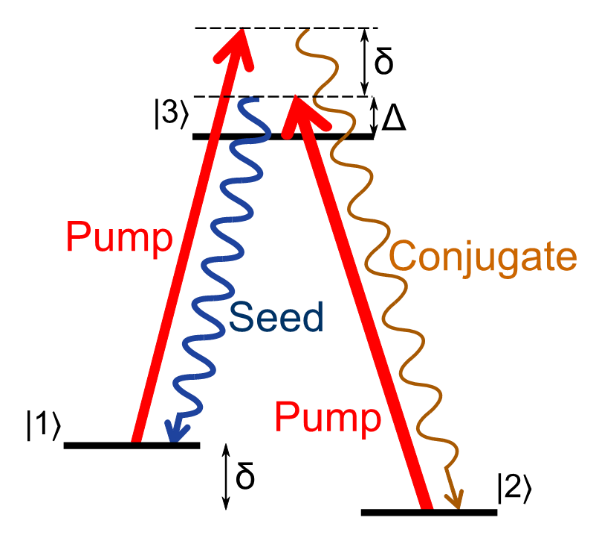}
\caption{The double-$\Lambda$ configuration for 4WM.}
\end{figure}

Here, the configuration is exploited when the frequencies of the pump and the seed reach a double-$\Lambda$ configuration with three atomic levels (see Fig.~1), so that when two pump photons are absorbed, the seed (signal) is amplified and a conjugate (idler) is generated. The three levels, in our case, are the ones associated with the $D_2$ line of Cesium: the hyperfine splitting $\delta$ of the ground state $6^{2}S_{1/2}$ are the levels $|1\rangle$ and $|2\rangle$ and the excited state $6^{2}P_{3/2}$ will be our level $|3\rangle$.
The effective Hamiltonian describing the generation of the twin beams is similar to the case of PDC and can be written as:
\begin{equation}
\mathcal{H}= \gamma [a_{1} a_ {2}+a^{\dag}_{1} a^{\dag}_{2}]\label{Hamiltonian}
\end{equation}
where $a_{i}$ is the annihilation operator of the optical mode $i=1,2$ of the twin beam, while the pump is considered as a classical field and determines the strength of the coupling ( $\gamma\propto \sqrt{I_{p}}$, $I_{p}$ being the pump intensity). In this paper we consider the stimulated process in which one mode is seeded with a coherent state (a laser) and the other one is in the vacuum state. In the spectral domain the 4WM process is efficient when
\begin{equation}
\omega_{conj}=\omega_{seed}+2\delta
\end{equation}
with $\omega_{conj}$ and $\omega_{seed}$ the frequency of the conjugate beam and the seed beam respectively.
The mean number of photon detected from the two beams are  $\langle N_1 \rangle=\eta\lambda (1+\mu)$ for the conjugate beam, and $\langle N_2 \rangle=\eta[\lambda+\mu (1+\lambda)]$  for the amplified seed beam, where $\mu$  is the number of photons per mode of the seed, $\lambda$ is the number of photons per mode of the spontaneous emission and $\eta$ is the detection efficiency (assumed equal in both channels). The non-classical correlations of the fields are witnessed by the fact that the variance on the difference of the two beams intensity falls below the shot noise limit, i.e. $V(N_{1}-N_{2})/\langle N_{1}+N_{2}\rangle = 1-\eta[1-1/(2g-1)]$, for $\mu\gg1$. Here $g$ represents the gain of the process defined as $g=N_{1}/(\eta\mu)\simeq 1+\lambda$.

\section{Experiment}

Our setup consists of (see Fig.~2) a 1 cm long glass cell, containing $^{133}$Cs (Cesium) vapor, maintained at about $110^{\circ}$C inside an oven with open windows. Two identical CW semi-conductor diode lasers with 852 nm wavelength (EYB-DFB-0852) are used for the pump and the seed. The frequency at which they emit is set precisely by adjusting their temperature and current with laser controllers (Thorlabs ITC 4001 for the seed, and ILX Lightwave LDC-3724C for the pump).
\begin{figure}
\centering \includegraphics[width=12cm]{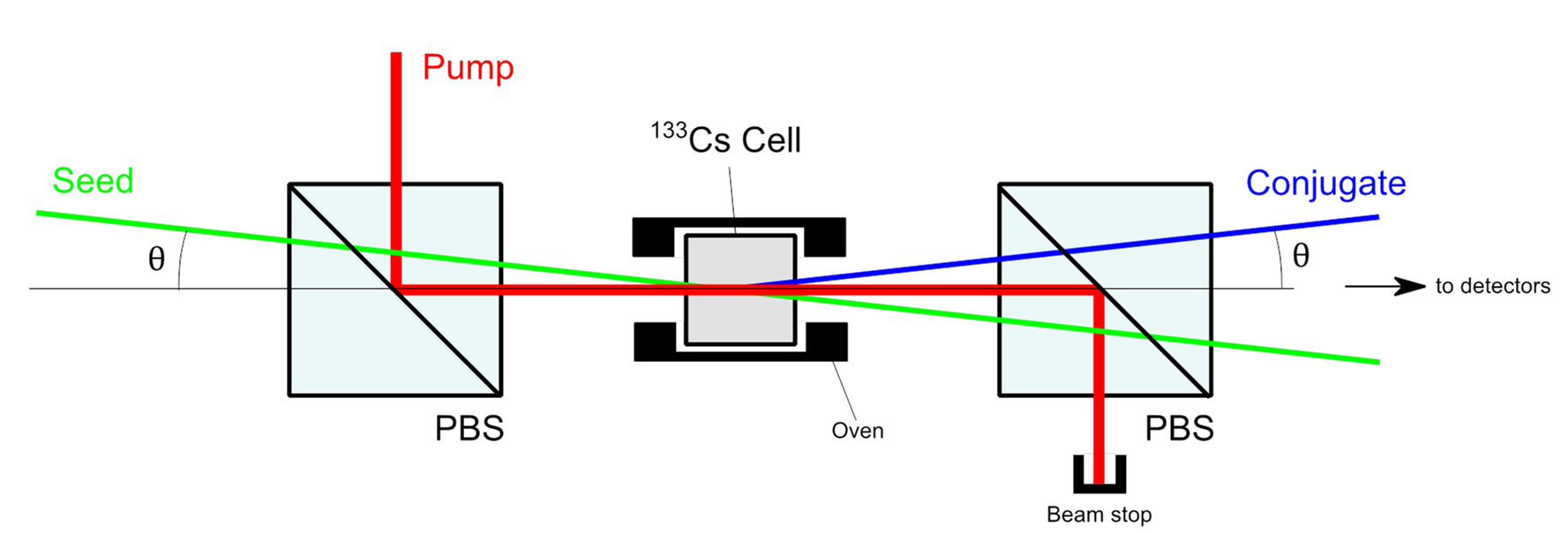}
\caption{Setup for 4WM in the non-collinear case: The pump (red) and the seed (green) are orthogonally polarized and overlap in a cell of Cesium vapor placed in an oven, thereby generating a conjugate beam (blue), together with an amplification of the seed (green). The pump is removed by its orthogonal polarization with the use of a PBS.}
\end{figure}
In order to obtain “clean” Gaussian beams for the seed and the pump, we couple each of them into polarization-maintaining single-mode fibers (PMJ-3A, F3AF-850-5/125 by OZ Optics)  While this cleaning of the spatial modes is not mandatory to observe the 4WM signature in the collinear case, it becomes quite important in the non-collinear case to get a clear distinction between the spatial modes of the three beams. Optical isolators (I-80-T4-M at 852 nm by Isowave) are used to avoid reflections backwards the diodes and  minimizing unwonted interference effects and instability of the lasers emission. It should be noted that the adjustment of the settings of the optical isolators is particularly sensitive in order to get a good 4WM signal.

A slightly incorrect setting in the optical isolator will result in an instable comb in the frequency domain that is quite detrimental to a good 4WM process. A Glan-Thompson Polarizer (with high Extinction Ratio of 100,000:1) is used at the exit of the cell to discard the pump (vertically polarized) while letting the seed and the conjugate (horizontally polarized) pass  towards the detectors. In the collinear configuration, the three co-propagating beams are focused on a photodiode, while when operating in the non-collinear configuration, the three spatially distinct beams are sent on a CCD camera (see Fig.~3).

\begin{figure}
\centering \includegraphics[width=12cm]{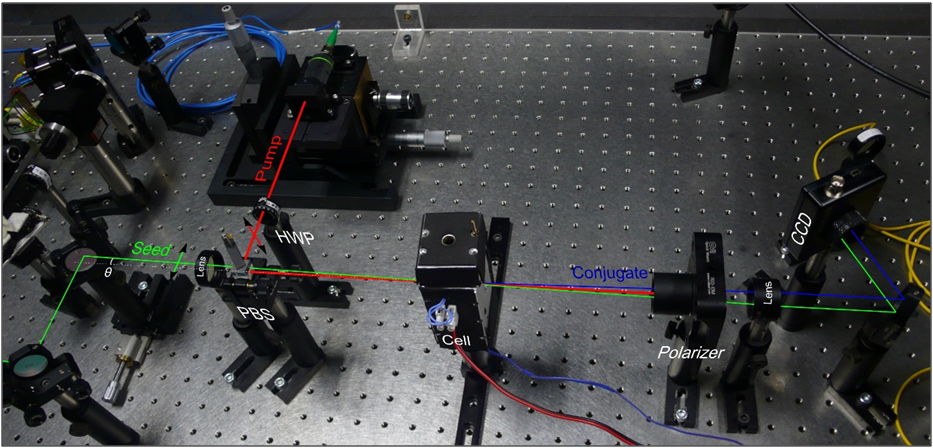}
\caption{Experimental setup for the non-collinear case. The (fake) colors of the three beams are meant only to distinguish them for clarity, but the three beams i.e the pump, the seed and the conjugate---have all a precisely set wavelength close to 852 nm, separated in frequency by only few GHz. The glass cell containing the cesium gas is not directly visible here, as it is placed inside an oven. The pump and the seed overlap inside the cell of Cesium vapor. The part of the setup devoted to “clean” the spatial modes of the diode laser (with single-mode fibers) is not shown here, and neither is the laser amplifier for the pump. In the collinear case, the seed is aligned with the pump at the output of the PBS, and the CCD camera is replaced by a photodiode upon which the entire signal is focused.}
\end{figure}

The frequency of the seed is then modulated by an external sweep function generator providing a triangular waveform (in current) around the D2 absorption line.	The pump is detuned from either of the absorption lines (negative or positive single-photon detuning $\Delta$), which allows to see the 4WM signal as an intensity peak (or in some cases dip) on the background of the absorption profile of the seed whenever frequency of the seed is at the correct 2-photon detuning $\delta$ to be in a double-$\Lambda$ configuration (see Fig.~4).
\begin{figure}
\centering \includegraphics[width=11cm]{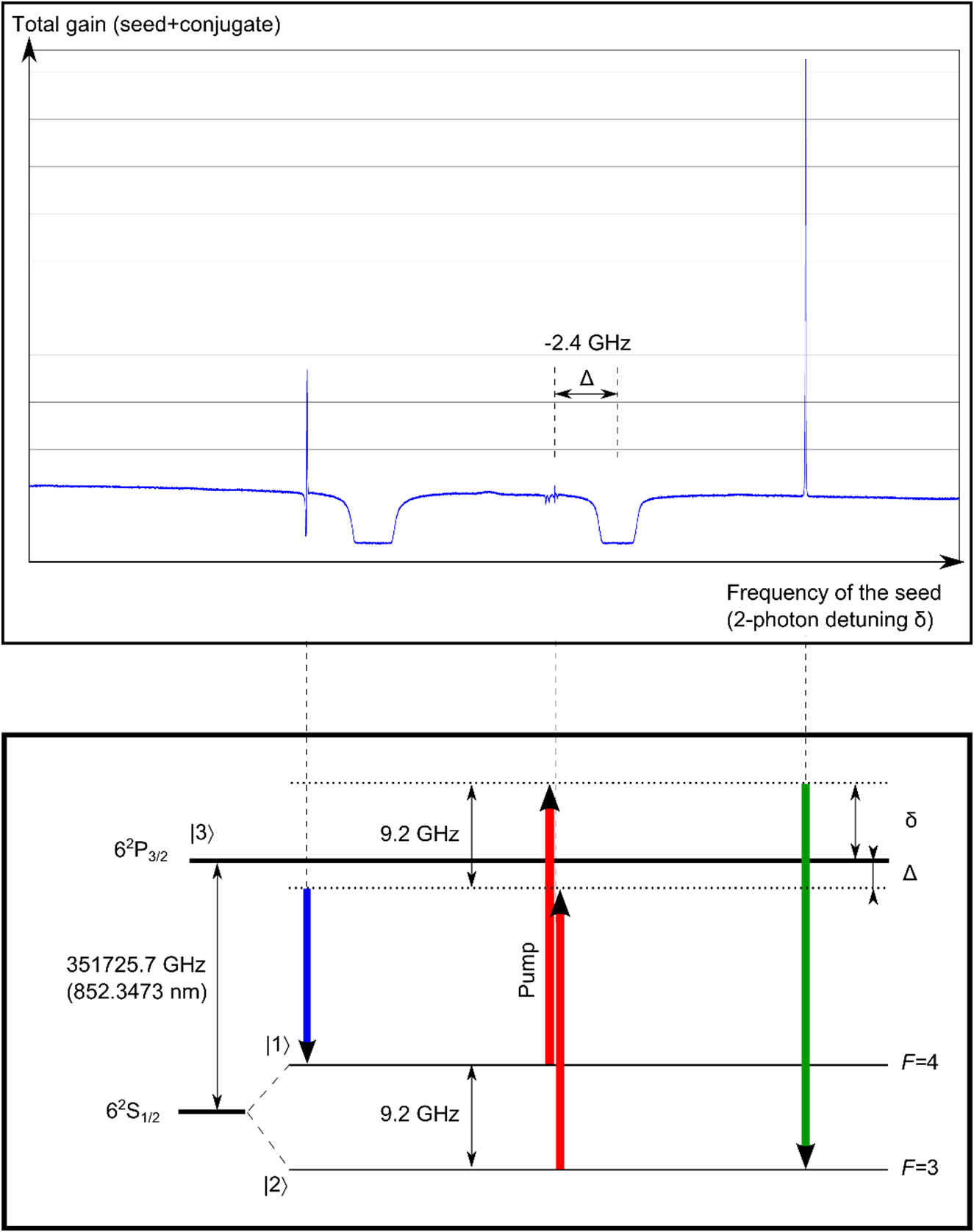}
\caption{The upper figure shows the highest measured gain in the collinear case, which includes both the gain on the seed and on the conjugate (so that the gain on the seed can be estimated to be $g\approx 5.6$). The lower figure shows the corresponding energy levels for the seed, pump and conjugate beams.}
\end{figure}

\section{Results}
In the collinear case, the three beams (the seed, the pump and the conjugate) are all propagating in the same direction, which maximizes the overlapping of the pump and the seed inside the cell, thus making for the highest possible gain with other parameters kept constant. This collinear configuration is well suited to start observing the 4WM phenomena and optimize the other parameters (such as the temperature of the cell, the frequency of the lasers, the precise overlapping of the beams, the adjustment of the optical isolators, etc). One can then gradually increase the angle of incidence in the cell between the pump and the laser towards a non-collinear configuration.
The pump was maintained at the highest achievable power of 110 mW. By contrast, the seed was kept at low power (less than 100 $\mu$W). Indeed, we observed the best gain with a power $<10$ $\mu$W for the seed. The seed beam and pump beam were focused on the cell, with a waist of 50 $\mu$m for the seed, and 200 $\mu$m for the pump.

For positive detuning on the $\vert 3\rangle \rightarrow \vert 2\rangle$ transition, the maximum gain $g\approx 3.1$ was obtained for a detuning of $\Delta=+1.8$ GHz (not shown here), at which the pump is rather heavily scattered (or possibly subjected to Kerr induced self-focusing) and too depolarized at the exit of the cell to be well removed.
The 4WM signal was then observed over a wide range of 1-photon detuning $\Delta$ from the $\vert 3\rangle \rightarrow\vert 2\rangle$ transition including negative detuning as well as detuning from the $\vert 3\rangle \rightarrow\vert 1\rangle$ transition (see Fig.~5).
For small detuning $\Delta$, the pump is depolarized at the exit of the cell, in particular for values close to the lowest energy transition $\vert 3\rangle \rightarrow\vert 1\rangle$.
The highest gain $g\approx 5.6$ was obtained for negative detuning on the $\vert 3\rangle \rightarrow\vert 2\rangle$ transition of $\Delta=-2.4$ GHz (see Fig.~4). With such a large detuning, the pump is less absorbed and depolarized. The spatial and polarization modes of the pump are better preserved and easier to discriminate from the seed and conjugate, both in the collinear and non-collinear configuration.

\begin{figure}
\centering \includegraphics[width=12cm]{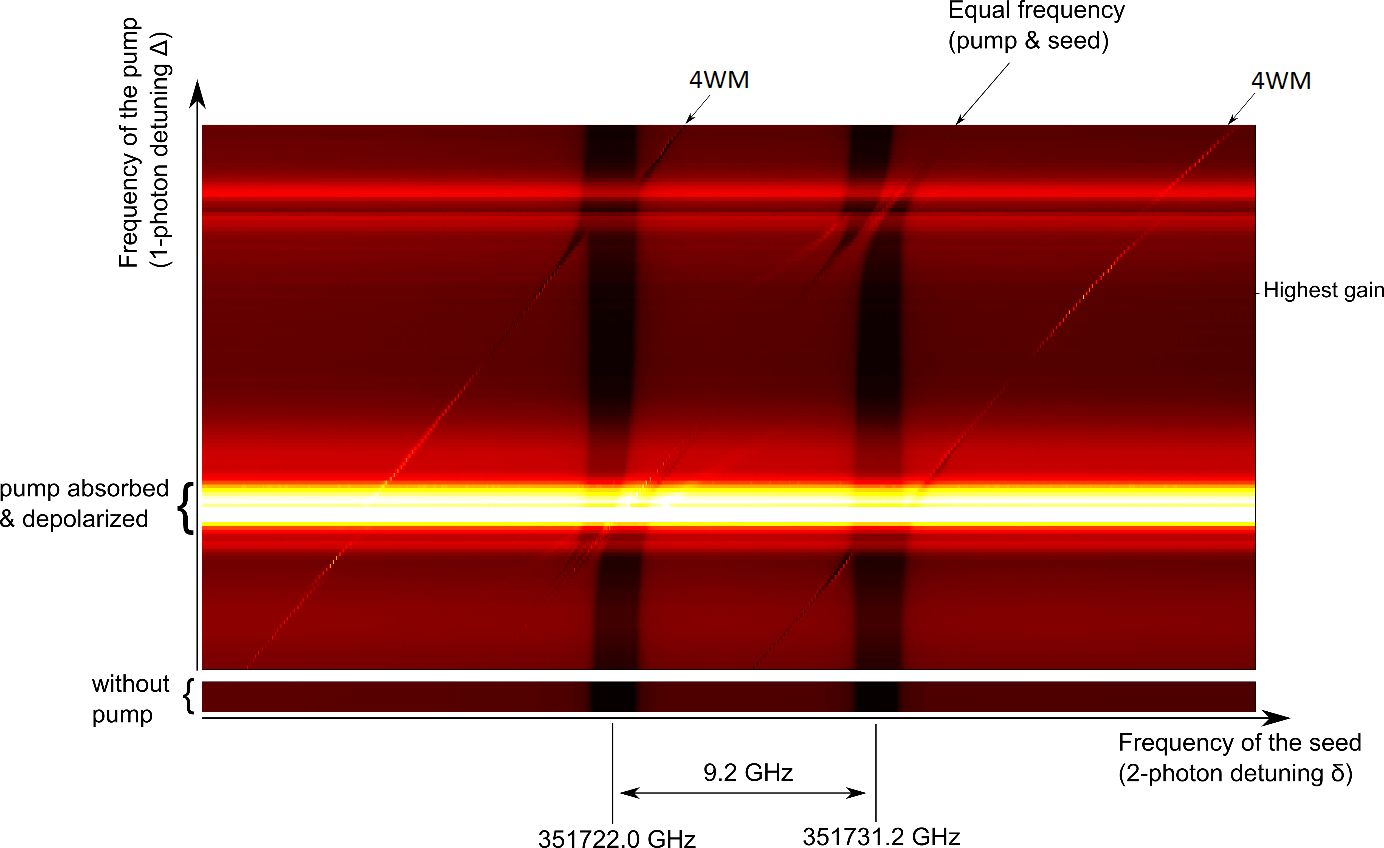}
\caption{Frequency map of the 4WM process. This image was obtained by collecting sweep graphs as in Fig.~4 for various single-photon detunings. Each horizontal line corresponds to a fixed frequency of the pump, whereas the frequency of the seed is swept through the absorption lines from $6^{2}P_{3/2}$ to the hyperfine levels of $6^{2}S_{1/2}$. The measured intensity at the photodiode for each sweep is then mapped to a color scale with arbitrary units (low intensity=dark; high intensity = bright).}
\end{figure}

An important effect that is quite noticeable on Fig.~5 is that the transitions to hyperfine levels are shifted by the intense pump. We think that this the consequence of light-shift effect (a.c. Stark shift). Consequently, the frequency difference between the seed and the conjugate is not always exactly twice the 9.2 GHz of the hyperfine splitting that one would expect in 4WM (we observed a variation of about 10$\%$ in the frequency difference between the seed and the conjugate, depending on the single-photon detuning).

As a further characterization of our source we studied the dependence on cell temperature. The optical thickness increases with the temperature inside the cell, and so does the gain of the 4WM (see Fig.~6). However, because the pump beam is close in frequency to the transitions to the hyperfine levels, it is partially absorbed, and therefore scattered and depolarized at the exit of the cell, possibly with a Kerr induced self-focusing of the pump beam. Increasing the optical thickness increases this effect (see Fig.~7), which is quite unwanted since we want to be able to distinguish and discard the pump both by its spatial mode and by it polarization mode. This effect can be seen in the frequency domain as a function of the single-photon detuning: it results in the horizontal bright bars on Fig.~5 when the pump is close to the atomic transitions, in particular the transition $\vert 3\rangle \rightarrow\vert 1\rangle$.

\begin{figure}
\centering \includegraphics[width=9cm]{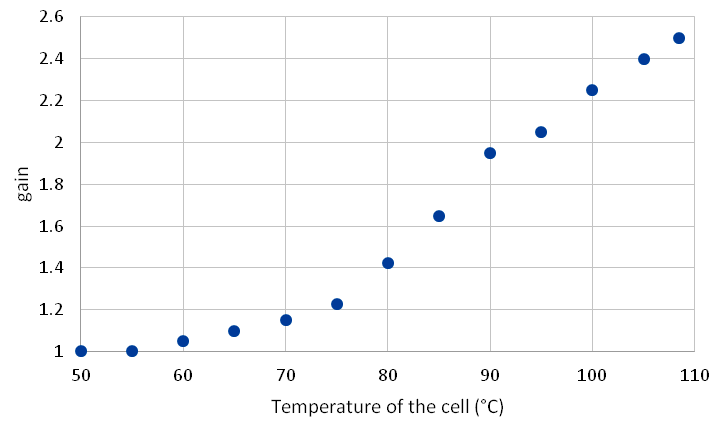}
\caption{Gain as a function of the temperature of the cell, for a single-photon detuning of $\Delta=+1.8$ GHz.}
\end{figure}

\begin{figure}
\centering \includegraphics[width=9cm]{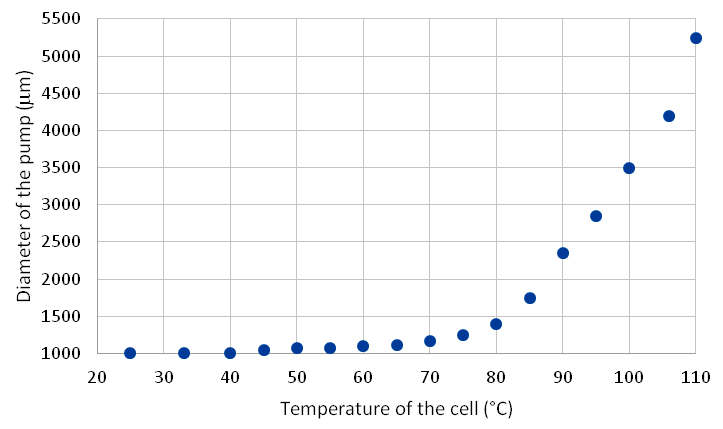}
\caption{Diameter of the pump beam 30 cm from the exit of the cell, as a function of the temperature of the cell for a single-photon detuning of $\Delta=+1.8$ GHz.This effect is clearly unwanted as it makes it difficult to separate the pump from the seed and the conjugate by their spatial modes.}
\end{figure}
The situation was much better for negative detuning $\vert 3\rangle \rightarrow\vert 2\rangle$, as the gain was not only higher, but the pump was as less affected by this unwanted effect.
Finally, we studied the dependence on pump power. In the region under study, the gain depends linearly on the pump power (see Fig.~8).
\begin{figure}
\centering \includegraphics[width=9cm]{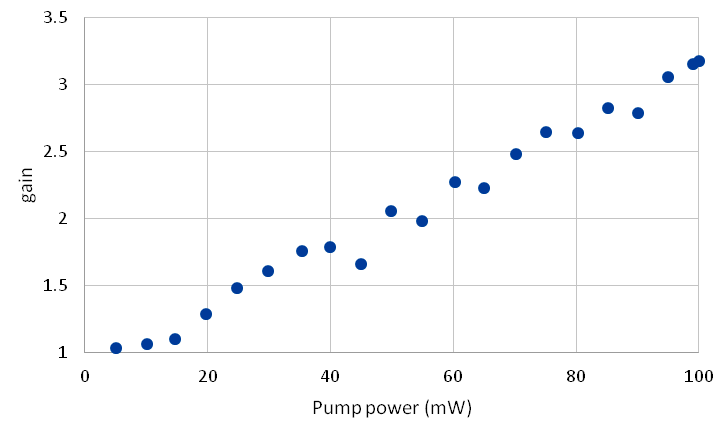}
\caption{Measured gain as a function of the pump power.}
\end{figure}

As a last step towards an exploitable source we considered the non-collinear case, where the seed is tilted with a small angle from the pump inside the cell, which allows to spatially distinguish the three output beams (see Figs. 2 and 3).
With the relatively large (and negative) single-photon detuning used here $\Delta=-2.4$ GHz, the pump retained well its spatial coherence and polarization (see Figs 9 and 10).
\begin{figure}
\centering \includegraphics[width=11cm]{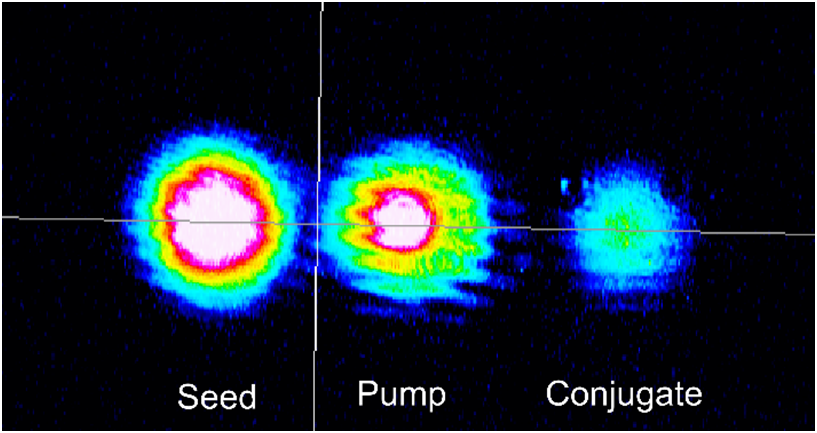}
\caption{Non-colienear 4WM with the pump allowed to partially pass through (for demonstration purpose).}
\end{figure}
By contrast, with second highest gain measured, with positive single-photon detuning $\Delta=+1.8$ GHz, the pump was much more scattered and depolarized at the exit of the cell (not shown here).

\begin{figure}
\centering \includegraphics[width=11cm]{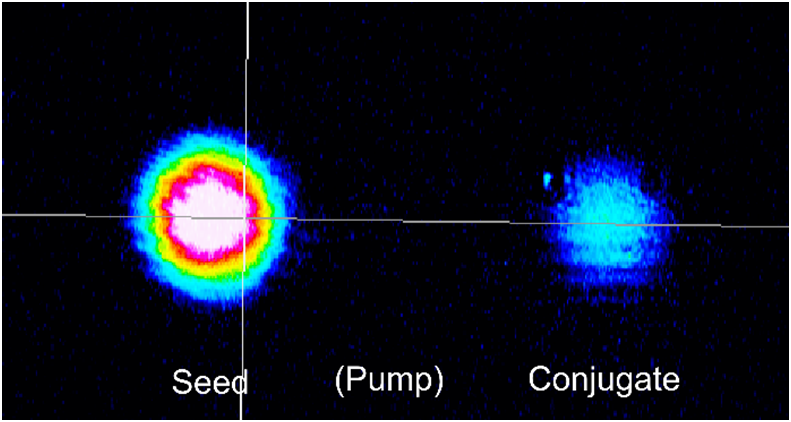}
\caption{Non-colinear 4WM with pump filtered by polarization.}
\end{figure}
\section{Conclusions}
We reported in details the realization of a first prototype of a twin beams source based on the $D_2$ line of Cesium atoms. Comprehensive study on the 4WM emission, function of the pump detuning with respect to the line of cesium, has been done. In particular, we achieved a gain $g\approx 5.6$ in collinear configuration by detuning the pump of  $\Delta=-2.4$ GHz with respect to $|3\rangle \rightarrow |2\rangle$ transition (see Fig.~4 and 5). A characterization of the gain of 4WM in function of the temperature of the cell (in the range of $20^{\circ} -110^{\circ}$ C), and  the pump power (in the range 10-100 mW) has been performed. The indication is that the gain of the source could be easily improved by increasing the pump intensity.

\section*{Acknowledgments}

We thank  Michael Pooser for useful discussions and precious suggestions.

\end{document}